\documentclass[12pt,a4paper]{article}

\title{Single rotation two-charge black holes in gauged supergravity}
\author{David D. K. Chow}
\date{}

\usepackage{amsmath}
\usepackage{amsfonts}
\usepackage{amssymb}
\usepackage{latexsym}

\makeatletter
\@addtoreset{equation}{section}
\makeatother

\newcommand{\be}{\begin{equation}}
\newcommand{\ee}{\end{equation}}
\newcommand{\ben}{\begin{equation}}
\newcommand{\een}{\end{equation}}
\newcommand{\bea}{\setlength\arraycolsep{2pt} \begin{eqnarray}}
\newcommand{\eea}{\end{eqnarray}}

\newcommand{\nnr}{\nonumber \\}
\newcommand{\lbl}{\label}

\newcommand{\eq}[1]{(\ref{#1})}
\newcommand{\se}{\section}
\newcommand{\sse}{\subsection}
\newcommand{\ssse}{\subsubsection}

\newcommand{\qd}{\quad}
\newcommand{\qqd}{\qquad}

\newcommand{\lt}{\left}
\newcommand{\rt}{\right}
\newcommand{\fr}{\frac}

\newcommand{\tf}{\tfrac}

\newcommand{\wtd}{\widetilde}

\newcommand{\df}{\textrm{d}}
\newcommand{\expe}[1]{\textrm{e}^{#1}}

\newcommand{\ol}{\overline}
\newcommand{\pd}{\partial}
\newcommand{\ra}{\rightarrow}

\newcommand{\sr}{\sqrt}

\newcommand{\grad}{\nabla}

\newcommand{\ga}{\alpha}
\newcommand{\gb}{\beta}

\newcommand{\gC}{\Gamma}
\newcommand{\gd}{\delta}
\newcommand{\gD}{\Delta}

\newcommand{\gq}{\theta}

\newcommand{\gvq}{\vartheta}

\newcommand{\gk}{\kappa}
\newcommand{\gl}{\lambda}
\newcommand{\gL}{\Lambda}

\newcommand{\gS}{\Sigma}

\newcommand{\gf}{\phi}
\newcommand{\gF}{\Phi}
\newcommand{\gvf}{\varphi}

\newcommand{\gw}{\omega}
\newcommand{\gW}{\Omega}

\newcommand{\uF}{\textrm{F}}

\newcommand{\uU}{\textrm{U}}

\newcommand{\im}{\textrm{i}}

\newcommand{\SO}{\textrm{SO}}

\newcommand{\SU}{\textrm{SU}}

\newcommand{\cA}{\mathcal{A}}

\newcommand{\cL}{\mathcal{L}}

\newcommand{\cN}{\mathcal{N}}
\newcommand{\cO}{\mathcal{O}}

\newcommand{\ds}{\textrm{d} s^2}
\newcommand{\dsi}{\left( \frac{\partial}{\partial s} \right) ^2}

\begin{document}


\thispagestyle{empty}

\begin{flushright}
MIFPA-11-37
\end{flushright}
\vspace*{100pt}
\begin{center}
\textbf{\Large{Single-rotation two-charge black holes in gauged supergravity}}\\
\vspace{50pt}
\large{David D. K. Chow}
\end{center}

\begin{center}
\textit{George P. \& Cynthia W. Mitchell Institute for Fundamental Physics \& Astronomy,\\
Texas A\&M University, College Station, TX 77843-4242, USA}\\
{\tt chow@physics.tamu.edu}\\
\vspace{30pt}
{\bf Abstract\\}
\end{center}
We consider asymptotically AdS, non-extremal, charged and rotating black holes with rotation in a single 2-plane and two independent $\uU (1)$ charge parameters.  Using a common ansatz, solutions are found for 5-dimensional $\uU (1)^3$ gauged supergravity, 7-dimensional $\uU(1)^2$ gauged supergravity, and 6-dimensional $\uU (1)$ gauged supergravity coupled to matter.  We also find static AdS black holes with two $\uU (1)$ charges of a certain theory in arbitrary dimensions.  Some basic properties of the solutions are studied.

\newpage


\se{Introduction}


The Myers--Perry metric \cite{myeper} is a solution of Einstein gravity without a cosmological constant in arbitrary spacetime dimension $D \geq 4$.  It represents an asymptotically flat black hole with the maximum number of independent rotation parameters.  Two generalizations of the most general Myers--Perry solution are known in arbitrary spacetime dimension $D \geq 4$.  One generalization is the Kerr--anti-de Sitter (AdS) metric \cite{gilupapo1, gilupapo2}, which allows for a cosmological constant, and represents an asymptotically AdS black hole.  The other generalization is the 2-charge Cveti\v{c}--Youm solution \cite{cveyou}, which allows for two $\uU(1)$ charges, and represents a charged asymptotically flat black hole.  It is of interest to find further AdS generalizations of these black hole solutions as gravitational backgrounds for studying the AdS/CFT correspondence.  Non-extreme AdS black holes are are of interest for understanding field theories at non-zero temperature.  Extreme black holes, supersymmetric or non-supersymmetric, are of interest as their entropy can be better understood.

In dimensions $D = 4, 5, 6, 7$, the 2-charge Cveti\v{c}--Youm solution gives solutions of ungauged supergravity.  The Kerr--AdS metric is a solution of Einstein gravity and hence a solution of gauged supergravity in these dimensions.  It should be possible to interpolate between these two solutions for $D = 4, 5, 6, 7$ in the context of gauged supergravity.  In $D = 6$, the two $\uU(1)$ gauge fields should be equal to recover a bosonic truncation of the Romans $\uF (4)$ gauged supergravity \cite{romans}.  In $D = 4, 5, 7$, the relevant theories are maximal gauged supergravities.  When the two $\uU (1)$ charges are equal, the interpolation is known and has a common structure for these dimensions \cite{chow2, chow1}.  More recently, the interpolation was achieved with two independent $\uU (1)$ charges in $D = 4$ \cite{chow3}, which is simpler because of the low dimensionality, after building on the insights of the simpler $D = 4$ solution with one $\uU (1)$ charge \cite{chow4}.

In this paper, we generalize the $D = 4$ solution to $D = 5, 6, 7$.  A simplification is to have rotation in only a single 2-plane.  This single-rotation simplification was used as an intermediate step in obtaining the general Myers--Perry metric \cite{myeper}, and gave the first examples of Kerr--AdS metrics in arbitrary dimension \cite{hahuta}.  We shall obtain here black hole solutions of gauged supergravity in $D = 5, 6, 7$ with a single non-zero rotation parameter and two independent and non-zero $\uU (1)$ charges.  The ansatz used is closely related to the $D = 4$ solution, and has the advantage of being well-adapted to the separability of the Hamilton--Jacobi and Klein--Gordon equations for these solutions.

The outline of this paper is as follows.  In Section 2, we review some relevant solutions in arbitrary dimensions that will guide us.  In Section 3, we obtain single-rotation 2-charge AdS black holes in dimensions $D = 5, 6, 7$.  In Section 4, we study properties of the solutions, namely thermodynamics, supersymmetry and hidden symmetries.  In Section 5, we specialize to non-rotating solutions, but generalize to a simple bosonic theory in arbitrary dimensions, finding 2-charge static AdS black holes.  We conclude in Section 6.

{\bf Note added:} While this paper was being completed, the papers \cite{wu2, wu3} appeared, which use a different ansatz to find the rotating solutions in $D = 5$ and $D = 7$.


\se{Solutions in arbitrary dimensions}


We first review two solutions that exist in arbitrary dimensions: the Kerr--AdS solution of Einstein gravity, and the Horowitz--Sen single-rotation black hole solution.  The single-rotation 2-charge AdS black hole solutions will interpolate between these.


\sse{Kerr-AdS metric}


The $D$-dimensional Kerr--AdS metric with rotation only in a single 2-plane was obtained in \cite{hahuta}, and can be expressed as
\bea
\ds & = & \fr{1}{(r^2 + y^2) \Xi^2} \bigg( - (V_y^2 \gD_r - V_r^2 \gD_y) \, \df t^2 + (\wtd{V}_r^2 \gD_y - \wtd{V}_y^2 \gD_r) a^2 \, \df \gf^2 - \fr{2 m \gD_y}{r^{D - 5} a} \, 2 \, \df t \, \df \gf \bigg) \nnr
&& + \fr{r^2 + y^2}{\gD_r} \, \df r^2 + \fr{r^2 + y^2}{\gD_y} \, \df y^2 + \fr{r^2 y^2}{a^2} \, \df \gW_{D - 4}^2 ,
\lbl{KerrAdS}
\eea
where
\bea
&& \gD_r = r^2 + a^2 - \fr{2 m}{r^{D - 5}} , \qd \gD_y = (1 - g^2 y^2) (a^2 - y^2) , \qd \Xi = 1 - a^2 g^2 , \nnr
&& V_r^2 = (1 + g^2 r^2)^2 , \qd V_y^2 = (1 - g^2 y^2)^2 , \qd \wtd{V}_r^2 = (1 + r^2 / a^2)^2 , \qd \wtd{V}_y^2 = (1 - y^2 / a^2)^2 ,
\eea
and $\df \gW_{D - 4}^2$ is the round metric on the unit sphere $S^{D - 4}$.  It is a solution of Einstein gravity with a cosmological constant, $G_{a b} + \gL g_{a b} = 0$, where $\gL = - \tf{1}{2} (D - 1) (D - 2) g^2$.  The coordinates here are asymptotically static, and $t$ and $\gf$ here are canonically normalized.  The solution in the original coordinates of \cite{hahuta} is obtained by letting $y = a \cos \gq$ and performing the coordinate shift $\gf \ra \gf + a g^2 t$.


\sse{Horowitz--Sen solution}


A $D$-dimensional Lagrangian that commonly arises within supergravity is
\ben
\cL_D = R \star 1 - \fr{1}{2} \sum_{i = 1}^2 \star \df \gvf_i \wedge \df \gvf_i - \fr{1}{2} \sum_{I = 1}^2 X_I^{-2} \star F_{(2)}^I \wedge F_{(2)}^I - \fr{1}{2} X_1^{-2} X_2^{-2} \star H_{(3)} \wedge H_{(3)} ,
\een
where the field strengths are given in terms of 1-form potentials $A_{(1)}^I$ and a 2-form potential $B_{(2)}$ by $F_{(2)}^I = \df A_{(1)}^I$ and
\ben
H_{(3)} = \df B_{(2)} - \tf{1}{2} A_{(1)}^1 \wedge F_{(2)}^2 - \tf{1}{2} A_{(1)}^2 \wedge F_{(2)}^1 ,
\een
and we have introduced the scalar combinations
\ben
X_1 = \expe{- \gvf_1 / \sr{2 (D - 2)} - \gvf_2 / \sr{2}} , \qd X_2 = \expe{- \gvf_1 / \sr{2 (D - 2)} + \gvf_2 / \sr{2}} .
\een
For example, when $4 \leq D \leq 9$, it is a consistent bosonic truncation of the reduction on a torus of heterotic supergravity.

It can also be convenient to dualize the 3-form field strength $H_{(3)}$ to a $(D - 3)$-form field strength $F_{(D - 3)} = \df A_{(D - 4)}$, via
\ben
F_{(D - 3)} = (-1)^D X_1^{-2} X_2^{-2} \star H_{(3)} .
\een
The field equations can be obtained from the dual Lagrangian
\bea
\cL_D & = & R \star 1 - \fr{1}{2} \sum_{i = 1}^2 \star \df \gvf_i \wedge \df \gvf_i - \fr{1}{2} \sum_{I = 1}^2 X_I^{-2} \star F_{(2)}^I \wedge F_{(2)}^I - \fr{1}{2} X_1^2 X_2^2 \star F_{(D - 3)} \wedge F_{(D - 3)} \nnr
&& - F_{(2)}^1 \wedge F_{(2)}^2 \wedge A_{(D - 4)} .
\lbl{Lagrangian}
\eea
Henceforth, we shall generally work with $F_{(D - 3)}$ rather than $H_{(3)}$.

The Horowitz--Sen solution \cite{horsen} solves this theory.  It represents a black hole rotating in a single 2-plane and carrying two $\uU (1)$ charges.  More generally, the 2-charge Cveti\v{c}--Youm solution \cite{cveyou} solves this theory, allowing for arbitrary rotation.  We use the notation of Cveti\v{c} and Youm, with parameters $\gd_1$ and $\gd_2$ that are related to the Horowitz and Sen parameters $\ga$ and $\gb$ by $\ga = \gd_1 + \gd_2$ and $\gb = \gd_1 - \gd_2$.  For rotation in only a single 2-plane, the metric is
\bea
\ds & = & \fr{(H_1 H_2)^{- (D - 3)/(D - 2)}}{r^2 + y^2} \bigg( - ( \gD_r - \gD_y) \, \df t^2 + ( \wtd{V}_r^2 \gD_y - \wtd{V}_y^2 \gD_r ) a^2 \, \df \gf^2 - \fr{2 m c_1 c_2 \gD_y}{r^{D - 5} a} \, 2 \, \df t \, \df \gf \bigg) \nnr
&& + (H_1 H_2)^{1 / (D - 2)} \bigg( \fr{r^2 + y^2}{\gD_r} \, \df r^2 + \fr{r^2 + y^2}{\gD_y} \, \df y^2 + \fr{r^2 y^2}{a^2} \, \df \gW_{D - 4}^2 \bigg) ,
\eea
and the other fields are given by
\bea
&& X_1 = H_1^{- (D - 1)/2 (D - 2)} H_2^{(D - 3)/2 (D - 2)} , \qd X_2 = H_1^{(D - 3)/2 (D - 2)} H_2^{- (D - 1)/2 (D - 2)} , \nnr
&& A_{(1)}^1 = \fr{2 m s_1}{H_1 (r^2 + y^2) r^{D - 5}} \lt( c_1 \, \df t - \fr{c_2 (a^2 - y^2)}{a} \, \df \gf \rt) , \nnr
&& A_{(1)}^2 = \fr{2 m s_2}{H_2 (r^2 + y^2) r^{D - 5}} \lt( c_2 \, \df t - \fr{c_1 (a^2 - y^2)}{a} \, \df \gf \rt) , \nnr
&& B_{(2)} = \lt( \fr{1}{H_1} + \fr{1}{H_2} \rt) \fr{m s_1 s_2 (a^2 - y^2)}{(r^2 + y^2) (r^2 + a^2) a} \, \df t \wedge \df \gf ,
\eea
where
\bea
&& \gD_r = r^2 + a^2 - \fr{2 m}{r^{D - 5}} , \qd \gD_y = a^2 - y^2 , \nnr
&& \wtd{V}_r^2 = \lt( 1 + \fr{r^2}{a^2} + \fr{2 m s_1^2}{r^{D - 5} a^2} \rt) \lt( 1 + \fr{r^2}{a^2} + \fr{2 m s_2^2}{r^{D - 5} a^2} \rt) , \qd \wtd{V}_y^2 = \lt( 1 - \fr{y^2}{a^2} \rt) ^2 , \nnr
&& H_I = 1 + \fr{2 m s_I^2}{(r^2 + y^2) r^{D - 5}} , \qd s_I = \sinh \gd_I , \qd c_I = \cosh \gd_I , \qd \Xi = 1 - a^2 g^2 .
\eea
We may take
\ben
A_{(D - 4)} = \fr{2 m s_1 s_2 y}{r^2 + y^2} \bigg( \fr{y}{a} \bigg) ^{D - 4} \gw_{(D - 4)} ,
\een
where $\gw_{(D - 4)}$ is the volume form on the unit sphere $S^{D - 4}$.  Our convention is that the $D$-dimensional volume form is
\ben
\star 1 = (H_1 H_2)^{1/(D - 2)} \fr{(r^2 + y^2)}{a} \bigg( \fr{r y}{a} \bigg) ^{D - 4} \, \df t \wedge \df r \wedge \df \gf \wedge \df y \wedge \gw_{(D - 4)} .
\een


\se{Black holes in gauged supergravities}


We now construct single-rotation 2-charge black holes in gauged supergravity in dimensions $D = 5, 6, 7$.


\sse{Gauged supergravities}


First, we review here the relevant gauged supergravity theories in $D = 5, 6, 7$.


\ssse{Five dimensions}


Reducing type IIB supergravity on $S^5$ leads to 5-dimensional maximal $\cN = 8$, $\SO (6)$ gauged supergravity.  It can be consistently truncated to $\uU (1)^3$ gauged supergravity, which is $\cN = 2$ gauged supergravity coupled to 2 vector multiplets.  The bosonic fields are a graviton, three $\uU (1)$ gauge fields and 2 dilatons.

The bosonic Lagrangian is
\bea
\cL_5 & = & R \star 1 - \fr{1}{2} \sum_{i = 1}^2 \star \df \gvf_i \wedge \df \gvf_i - \fr{1}{2} \sum_{I = 1}^3 X_I^{-2} \star F_{(2)}^I \wedge F_{(2)}^I - F_{(2)}^1 \wedge F_{(2)}^2 \wedge A_{(1)}^3 + 4 g^2 \sum_{I = 1}^3 X_I^{-1} \star 1 , \nnr
\eea
where $F_{(2)}^I = \df A_{(1)}^I$ and
\ben
X_1 = \expe{- \gvf_1 / \sr{6} - \gvf_2 / \sr{2}} , \qd X_2 = \expe{- \gvf_1 / \sr{6} + \gvf_2 / \sr{2}} , \qd X_3 = \expe{2 \gvf_1 / \sr{6}} .
\een
For the ungauged theory, with $g = 0$, this is the Lagrangian of \eq{Lagrangian}, with the third $\uU (1)$ gauge field being $A_{(1)}^3 = A_{(D - 4)}$.  The resulting field equations are
\bea
G_{a b} & = & \sum_{i = 1}^2 \bigg( \fr{1}{2} \grad_a \gvf_i \, \grad_b \gvf_i - \fr{1}{4} \grad^c \gvf_i \, \grad_c \gvf_i \, g_{a b} \bigg) + \sum_{I = 1}^3 X_I^{-2} \bigg( \fr{1}{2} F{^I}{_a}{^c} F{^I}{_{b c}} - \fr{1}{8} F^{I c d} F{^I}{_{c d}} g_{a b} \bigg) \nnr
&& + 2 g^2 \sum_{I = 1}^3 X_I^{-1} g_{a b} ,
\eea
and 
\bea
&& \square \gvf_1 = \fr{1}{2 \sr{6}} (X_1^{-2} F^{1 a b} F{^1}{_{a b}} + X_2^{-2} F^{2 a b} F{^2}{_{a b}} - 2 X_3^{-2} F^{3 a b} F{^3}{_{a b}}) - \fr{4}{\sr{6}} g^2 (X_1^{-1} + X_2^{-1} - 2 X_3^{-1}) , \nnr
&& \square \gvf_2 = \fr{1}{2 \sr{2}} (X_1^{-2} F^{1 a b} F{^1}{_{a b}} - X_2^{-2} F^{2 a b} F{^2}{_{a b}}) - 2 \sr{2} g^2 (X_1^{-1} - X_2^{-1}) , \nnr
&& \df (X_1^{-2} \star F_{(2)}^1) + F_{(2)}^2 \wedge F_{(2)}^3 = 0 , \nnr
&& \df (X_2^{-2} \star F_{(2)}^2) + F_{(2)}^3 \wedge F_{(2)}^1 = 0 , \nnr
&& \df (X_3^{-2} \star F_{(2)}^3) + F_{(2)}^1 \wedge F_{(2)}^2 = 0 .
\eea


\ssse{Six dimensions}


Reducing massive type IIA supergravity on $S^4$ leads to 6-dimensional $\cN = 4$, $\SU (2)$ gauged supergravity.  This has half of the maximum number of supersymmetries in 6 dimensions.  The theory can presumably be coupled to $\cN = 4$ matter, as discussed in \cite{cvgulupo}.

By analogy to other dimensions, we consider the bosonic Lagrangian
\bea
\cL_6 & = & R \star 1 - \fr{1}{2} \sum_{i = 1}^2 \star \df \gvf_i \wedge \df \gvf_i - \fr{1}{2} \sum_{I = 1}^2 X_I^{-2} \star F_{(2)}^I \wedge F_{(2)}^I - \fr{1}{2} X_1^2 X_2^2 \star F_{(3)} \wedge F_{(3)} \nnr
&& - F_{(2)}^1 \wedge F_{(2)}^2 \wedge A_{(2)} - g^2 X_1^{-1} X_2^{-1} \star A_{(2)} \wedge A_{(2)} - \fr{g^2}{3} A_{(2)} \wedge A_{(2)} \wedge A_{(2)} \nnr
&& + g^2 (9 X_1 X_2 + 6 X_1^{-3/2} X_2^{-1/2} + 6 X_1^{-1/2} X_2^{-3/2} - X_1^{-3} X_2^{-3}) \star 1 ,
\eea
where $F_{(2)}^I = \df A_{(1)}^I$ and
\ben
X_1 = \expe{- \gvf_1 / \sr{8} - \gvf_2 / \sr{2}} , \qd X_2 = \expe{- \gvf_1 / \sr{8} + \gvf_2 / \sr{2}} .
\een
If $A_{(1)}^1 = A_{(1)}^2 = A_{(1)}$ and $\gvf_2 = 0$, so $X_1 = X_2 = X$, then the bosonic theory is that of $\cN = 4$, $\SU (2)$ gauged supergravity, truncated to exciting only a $\uU (1)$ gauge field of the full $\SU (2)$ gauge group, studied, for example, in \cite{chow1}.  If we instead truncate the theory by taking $A_{(2)} = 0$, then we obtain a truncation of the theory in \cite{chlupo4}, in their notation $A_{(1)}^i = 0$ and $\gvf_i = 0$, rescaling $g \ra \tf{3}{2} g$.  The resulting field equations are
\bea
G_{a b} & = & \sum_{i = 1}^2 \bigg( \fr{1}{2} \grad_a \gvf_i \, \grad_b \gvf_i - \fr{1}{4} \grad^c \gvf_i \, \grad_c \gvf_i \, g_{a b} \bigg) + \sum_{I = 1}^2 X_I^{-2} \bigg( \fr{1}{2} F{^I}{_a}{^c} F{^I}{_{b c}} - \fr{1}{8} F^{I c d} F{^I}{_{c d}} g_{a b} \bigg) \nnr
&& + X_1^2 X_2^2 \bigg( \fr{1}{4} F{_a}{^{c d}} F_{b c d} - \fr{1}{24} F^{c d e} F_{c d e} g_{a b} \bigg) + 2 g^2 X_1^{-1} X_2^{-1} \bigg( \fr{1}{2} A{_a}{^c} A_{b c} - \fr{1}{8} A^{c d} A_{c d} g_{a b} \bigg) \nnr
&& + \fr{1}{2} g^2 (9 X_1 X_2 + 6 X_1^{-3/2} X_2^{-1/2} + 6 X_1^{-1/2} X_2^{-3/2} - X_1^{-3} X_2^{-3}) g_{a b} ,
\eea
and
\bea
&& \square \gvf_1 = \fr{1}{4 \sr{2}} \sum_{I = 1}^2 X_I^{-2} F^{I a b} F{^I}{_{a b}} - \fr{1}{6 \sr{2}} X_1^2 X_2^2 F^{a b c} F_{a b c} + \fr{1}{2 \sr{2}} g^2 X_1^{-1} X_2^{-1} A^{a b} A_{a b} \nnr
&& \qqd \qqd + \fr{3}{\sr{2}} g^2 (3 X_1 X_2 - 2 X_1^{-3/2} X_2^{-1/2} - 2 X_1^{-1/2} X_2^{-3/2} + X_1^{-3} X_2^{-3}) , \nnr
&& \square \gvf_2 = \fr{1}{2 \sr{2}} (X_1^{-2} F^{1 a b} F{^1}{_{a b}} - X_2^{-2} F^{2 a b} F{^2}{_{a b}}) + 3 \sr{2} g^2 (X_1^{-1/2} X_2^{-3/2} - X_1^{-3/2} X_2^{-1/2}) , \nnr
&& \df (X_1^{-2} \star F_{(2)}^1) + F_{(2)}^2 \wedge F_{(3)} = 0 , \nnr
&& \df (X_2^{-2} \star F_{(2)}^2) + F_{(2)}^1 \wedge F_{(3)} = 0 , \nnr
&& \df (X_1^2 X_2^2 \star F_{(3)}) + F_{(2)}^1 \wedge F_{(2)}^2 + 2 g^2 X_1^{-1} X_2^{-1} \star A_{(2)} + g^2 A_{(2)} \wedge A_{(2)} = 0 .
\eea


\ssse{Seven dimensions}


Reducing 11-dimensional supergravity on $S^4$ leads to 7-dimensional
maximal $\cN = 4$, $\SO (5)$ gauged supergravity.  It can be consistently truncated to $\uU (1)^2$ gauged supergravity, which is $\cN = 2$ gauged supergravity coupled to a vector multiplet.  The bosonic fields are a graviton, a 3-form potential, two $\uU (1)$ gauge fields and two scalars.

The bosonic Lagrangian is
\bea
\cL_7 & = & R \star 1 - \fr{1}{2} \sum_{i=1}^2 \star \df \gvf_i \wedge \df \gvf_i - \fr{1}{2} \sum_{I = 1}^2 X_I^{-2} \star F_{(2)}^I
\wedge F_{(2)}^I - \fr{1}{2} X_1^2 X_2^2 \star F_{(4)} \wedge
F_{(4)} \nnr
&& - F_{(2)}^1 \wedge F_{(2)}^2 \wedge
A_{(3)} - g F_{(4)} \wedge A_{(3)} \nnr
&& + 2 g^2 (8 X_1 X_2 + 4 X_1^{-1} X_2^{-2} + 4 X_1^{-2} X_2^{-1} -
X_1^{-4} X_2^{-4}) \star 1 ,
\lbl{Lagrangian7}
\eea
where $F_{(2)}^I = \df A_{(1)}^I$, $F_{(4)} = \df A_{(3)}$, and
\ben
X_1 = \expe{- \gvf_1 / \sr{10} - \gvf_2 / \sr{2}} , \qd X_2 =
\expe{ - \gvf_1 / \sr{10} + \gvf_2 / \sr{2}} .
\een
The resulting field equations are
\bea
G_{ab} & = & \sum_{i = 1}^2 \left( \fr{1}{2} \grad_a \gvf_i
\grad_b \gvf_i - \fr{1}{4} \grad^c \gvf_i \grad_c \gvf_i
g_{a b} \right) + \sum_{I = 1}^2 X_I^{-2} \left( \fr{1}{2} F{^I}{_{a}}{^c} F{^I}{_{b c}} - \fr{1}{8} F^{I c d} F{^I}{_{c d}} g_{a b} \right) \nnr
&& + X_1^2 X_2^2 \bigg( \fr{1}{12} F{_a}{^{c d e}} F_{b c d e} - \fr{1}{96} F^{c d e f} F_{c d e f} g_{a b} \bigg) \nnr
&& + g^2 (8 X_1 X_2 + 4 X_1^{-1} X_2^{-2} + 4
X_1^{-2} X_2^{-1} - X_1^{-4} X_2^{-4}) g_{a b} ,
\eea
and
\bea
&& \square \gvf_1 = \fr{1}{2 \sr{10}} \sum_{I = 1}^2 X_I^{-2} F^{I a b} F{^I}{_{a b}} - \fr{1}{12 \sr{10}} X_1^2 X_2^2 F^{a b c d} F_{a b c d} , \nnr
&& \qqd \qd + \fr{8}{\sr{10}} g^2 (4 X_1 X_2 - 3 X_1^{-1} X_2^{-2} - 3 X_1^{-2} X_2^{-1} + 2 X_1^{-4} X_2^{-4}) , \nnr
&& \square \gvf_2 = \fr{1}{2 \sr{2}} (X_1^{-2} F^{1 a b} F{^1}{_{a b}} - X_2^{-2} F^{2 a b} F{^2}{_{a b}}) + 4 \sr{2} g^2 (X_1^{-1} X_2^{-2} - X_1^{-2} X_2^{-1}) , \nnr
&& \df (X_1^{-2} \star F_{(2)}^1) + F_{(2)}^2 \wedge F_{(4)} = 0 , \nnr
&& \df (X_2^{-2} \star F_{(2)}^2) + F_{(2)}^1 \wedge F_{(4)} = 0 , \nnr
&& \df (X_1^2 X_2^2 \star F_{(4)}) + F_{(2)}^1 \wedge F_{(2)}^2 + 2 g F_{(4)} = 0 .
\eea
As well as the above field equations obtainable from the Lagrangian \eq{Lagrangian7}, there is also a self-duality condition to impose.  It can be stated by including a two-form potential $B_{(2)}$ and defining
\ben
H_{(3)} = \df B_{(2)} - \tf{1}{2} A_{(1)}^1 \wedge F_{(2)}^2 - \tf{1}{2} A_{(1)}^2 \wedge F_{(2)}^1 .
\een
The self-duality equation is
\ben
X_1^2 X_2^2 \star F_{(4)} = - 2 g A_{(3)} - H_{(3)} .
\een


\sse{Black hole solutions}


We now present the single-rotation black hole solutions of the above theories in $D = 5, 6, 7$.  The metrics are
\bea
\ds & = & \fr{(H_1 H_2)^{- (D - 3)/(D - 2)}}{(r^2 + y^2) \Xi^2} \bigg( - (V_y^2 \gD_r - V_r^2 \gD_y) \, \df t^2 + (\wtd{V}_r^2 \gD_y - \wtd{V}_y^2 \gD_r) a^2 \, \df \gf^2 \nnr
&& - \fr{2 m c_1 c_2 \wtd{c}_1 \wtd{c}_2 \gD_y}{r^{D - 5} a} \, 2 \, \df t \, \df \gf \bigg) \nnr
&& + (H_1 H_2)^{1/(D - 2)} \bigg( \fr{r^2 + y^2}{\gD_r} \, \df r^2 + \fr{r^2 + y^2}{\gD_y} \, \df y^2 + \fr{r^2 y^2}{a^2} \, \df \gW_{D - 4}^2 \bigg) , \lbl{metric}
\eea
and there are two scalars and two $\uU (1)$ gauge fields given by
\bea
&& X_1 = H_1^{- (D - 1)/2 (D - 2)} H_2^{(D - 3)/2 (D - 2)} , \qd X_2 = H_1^{(D - 3)/2 (D - 2)} H_2^{- (D - 1)/2 (D - 2)} , \nnr
&& A_{(1)}^1 = \fr{2 m s_1}{H_1 (r^2 + y^2) r^{D - 5} \Xi} \lt( c_1 (1 - g^2 y^2) \, \df t - \fr{c_2 (a^2 - y^2)}{a} \, \df \gf \rt) , \nnr
&& A_{(1)}^2 = \fr{2 m s_2}{H_2 (r^2 + y^2) r^{D - 5} \Xi} \lt( c_2 (1 - g^2 y^2) \, \df t - \fr{c_1 (a^2 - y^2)}{a} \, \df \gf \rt) ,
\eea
where
\bea
&& \gD_r = r^2 + a^2 - \fr{2 m}{r^{D - 5}} + g^2 \lt( r^2 + \fr{2 m s_1^2}{r^{D - 5}} \rt) \lt( r^2 + \fr{2 m s_2^2}{r^{D - 5}} \rt) + a^2 g^2 r^2 - \fr{2 m a^2 g^2 s_1^2 s_2^2}{r^{D - 5}} , \nnr
&& \gD_y = (1 - g^2 y^2) (a^2 - y^2) , \nnr
&& V_r^2 = \lt( 1 + g^2 r^2 + \fr{2 m s_1^2 g^2}{r^{D - 5}} \rt) \lt( 1 + g^2 r^2 + \fr{2 m s_2^2 g^2}{r^{D - 5}} \rt) , \qd V_y^2 = (1 - g^2 y^2)^2 , \nnr
&& \wtd{V}_r^2 = \lt( 1 + \fr{r^2}{a^2} + \fr{2 m s_1^2}{r^{D - 5} a^2} \rt) \lt( 1 + \fr{r^2}{a^2} + \fr{2 m s_2^2}{r^{D - 5} a^2} \rt) , \qd \wtd{V}_y^2 = \lt( 1 - \fr{y^2}{a^2} \rt) ^2 , \nnr
&& H_I = 1 + \fr{2 m s_I^2}{(r^2 + y^2) r^{D - 5}} , \qd s_I = \sinh \gd_I , \qd c_I = \cosh \gd_I , \qd \wtd{c}_I = \sr{1 + a^2 g^2 s_I^2} , \nnr
&& \Xi = 1 - a^2 g^2 .
\eea
There are slight differences regarding the $(D - 4)$-form potential $A_{(D - 4)}$.  In $D = 5$, there is a third $\uU (1)$ gauge field
\ben
A_{(1)}^3 = \fr{2 m s_1 s_2 y^2}{(r^2 + y^2) a} \gw_{(1)} ,
\een
In $D = 6$, there is a 2-form potential
\ben
A_{(2)} = \fr{2 m s_1 s_2 y^3}{(r^2 + y^2) a^2} \gw_{(2)} .
\een
In $D = 7$, there is a 3-form potential
\ben
A_{(3)} = \fr{2 m s_1 s_2 y^4}{(r^2 + y^2) a^3} \gw_{(3)} + \fr{2 m s_1 s_2 g y}{\Xi a r^2} \, \df t \wedge \df \gf \wedge \df y .
\een
The 2-form potential that enters the $D = 7$ self-duality equation, which more generally appears if one chooses to dualize $F_{(D - 3)}$ in favour of $H_{(3)}$, is
\ben
B_{(2)} = \lt( \fr{1}{H_1} + \fr{1}{H_2} \rt) \fr{m s_1 s_2 (1 - g^2 y^2) (a^2 - y^2)}{(r^2 + y^2) (r^2 + a^2) a \Xi} \, \df t \wedge \df \gf .
\een
For definiteness, we can introduce explicit coordinates for the round sphere $S^{D - 4}$, taking
\ben
\df \gW_1^2 = \df \gf_2^2 , \qd \df \gW_2^2 = \df \gvq^2 + \sin^2 \gvq \, \df \gf_2^2 , \qd \df \gW_3^2 = \df \gvq^2 + \sin^2 \gvq \, \df \gf_2^2 + \cos^2 \gvq \, \df \gf_3^2 .
\een
The form of the solutions here is valid also for $D = 4$ \cite{chow3}.

We have presented the solution using asymptotically static Boyer--Lindquist-type coordinates with canonical normalization.  Letting $y = a \cos \gq$ recovers the usual latitudinal coordinate $\gq$.  If $m = 0$, then the coordinate change
\ben
\Xi \hat{r}^2 \sin^2 \hat{\gq} = (r^2 + a^2) \sin^2 \gq , \qd \hat{r}^2 \cos^2 \hat{\gq} = r^2 \cos^2 \gq ,
\een
gives anti-de Sitter spacetime in the canonical form
\ben
\ds = - (1 + g^2 \hat{r}^2) \, \df t^2 + \fr{\df \hat{r}^2}{1 + g^2 \hat{r}^2} + \hat{r}^2 \, \df \gW_{D - 2}^2 ,
\een
where
\ben
\df \gW_{D - 2}^2 = \df \hat{\gq}^2 + \sin^2 \hat{\gq} \, \df \gf^2 + \cos^2 \hat{\gq} \, \df \gW_{D - 4}^2
\een
is the round metric on $S^{D - 2}$.

The solution is invariant under the discrete inversion symmetry
\ben
a \ra \fr{1}{a g^2} , \qd r \ra \fr{r}{a g} , \qd y \ra \fr{y}{a g} , \qd m \ra \fr{m}{(a g)^{D - 1}} , \qd \gf \ra g t , \qd t \ra \fr{\gf}{g} , \qd s_I \ra a g s_I .
\een
This symmetry is possessed by the Kerr--AdS metrics \cite{chlupo3, chlupo2} and by the 2-charge black hole solution of 4-dimensional $\uU (1)^4$ gauged supergravity \cite{chow4, chow3}.

The solutions carry 5 parameters: a mass parameter $m$, a rotation parameter $a$, two charge parameters $\gd_1$ and $\gd_2$, and a gauge-coupling constant $g$.Without charge, when $\gd_1 = \gd_2 = 0$, the solution reduces to the single-rotation Kerr--AdS metrics.  Without gauging, when $g = 0$, the solution reduces to the Horowitz--Sen solution.  The ungauged solutions can also be thought of as the 2-charge Cveti\v{c}--Youm solution \cite{cveyou} (or in $D = 5$ the 3-charge Cveti\v{c}--Youm solution \cite{cveyou2} with 2 non-zero charge parameters), specialized to rotation in a single 2-plane.  Without rotation, when $a = 0$, the solution reduces to static solutions in $D = 5$ \cite{becvsa}, $D = 6$ (for equal charges $\gd_1 = \gd_2$) \cite{cvlupo}, and $D = 7$ \cite{cvegub, liumin}.  With both charges equal, when $\gd_1 = \gd_2$, the solution reduces to solutions in $D = 5$ \cite{chcvlupo1}, $D = 6$ \cite{chow1}, $D = 7$ \cite{chow2}, specialized to rotation in a single 2-plane; to match the previous literature, one must change parameters, like in $D = 4$ \cite{chow3}.  With one non-zero charge, when $\gd_2 = 0$, the solution to 1-charge solutions in $D = 5$ \cite{chcvlupo2} and in general dimensions \cite{wu}, specialized to rotation in a single 2-plane.

The $D = 5$ solution can be lifted to type IIB supergravity in an explicit manner \cite{cveticetal}.  If the two charges are equal, then the $D = 6$ solution can be lifted to massive type IIA supergravity.  If there is a single non-zero $\uU (1)$ charge, then the $D = 7$ solution can be explicitly lifted to 11-dimensional supergravity \cite{cveticetal}; otherwise, the explicit lifting is not known.


\se{Properties of the rotating black holes}


We now study various properties of the single-rotation black holes, such as thermodynamics, supersymmetric solutions and hidden symmetries.  Because solutions in different dimensions are similar, we can present the results for $D = 4, 5, 6, 7$ together in a unified manner.  The results for $D = 4$ appeared in \cite{chow3}.


\sse{Thermodynamics}


For a black hole, $\gD_r (r)$ has a positive root at the outer horizon, say at $r = r_+$.  The angular velocity $\gW$ is constant over the horizon and is obtained from the Killing vector
\ben
l = \fr{\pd}{\pd t} + \gW \, \fr{\pd}{\pd \gf}
\een
that becomes null on the horizon.  The electrostatic potentials $\gF_I$ and surface gravity $\gk$ are also constant over the horizon.  They are given by evaluating on the horizon both $\gF_I = l \cdot A_{(1)}^I$ and $l^b \grad_b l^a = \gk l^a$, and then the Hawking temperature is $T = \gk / 2 \pi$.  The angular momentum and electric charges are given by the respective integrals
\ben
J = \fr{1}{16 \pi} \int_{S_\infty^{D - 2}} \! \star \df K , \qd Q_I = \fr{1}{16 \pi} \int_{S_\infty^{D - 2}} \! X_I^{-2} \star F_{(2)}^I + \ldots ,
\een
where $K = K_a \, \df x^a$ is the 1-form obtained from the Killing vector $K^a \, \pd_a = \pd / \pd \gf_1$, and the ellipsis represents $D$-dependent terms from the $A_{(1)}^I$ field equation, which for our solutions have zero contribution to $Q_I$.  The Bekenstein--Hawking entropy is $S = A/4$, where $A$ is the horizon area.

One finds that $T \, \df S + \gW \, \df J + \gF_1 \, \df Q_1 + \gF_2 \, \df Q_2$ is an exact differential, and so we may integrate the first law of black hole mechanics, $\df E = T \, \df S + \gW \, \df J + \gF_1 \, \df Q_1 + \gF_2 \, \df Q_2$, to obtain the thermodynamic mass $E$.  The same $E$ is obtained by calculating the AMD mass \cite{ashdas}.  For this definition, one introduces a metric $\ol{g}_{a b} = \gW^2 g_{a b}$ with both $\gW = 0$ and $\df \gW \neq 0$ on the conformal boundary.  The AMD mass in $D$ dimensions is \cite{ashdas}
\ben
E = \fr{1}{8 (D - 3) \pi g^3} \int_\gS \! \df \ol{\gS}_a \, \gW^{- (D - 3)} \ol{n}^c \ol{n}^d \ol{C}{^a}{_{c b d}} K^b ,
\een
where $\df \ol{\gS}_a$ is the area element of the $S^{D - 2}$ section of the conformal boundary, $K^a \, \pd_a = \pd / \pd t$, $\ol{n}_a = \pd_a \gW$ and $\ol{C}{^a}{_{b c d}}$ is the Weyl tensor.  For definiteness, we take $\gW = 1 / g r$.  One needs the behaviour of $C{^t}{_{r t r}}$ as $r \ra \infty$ for our solutions, which is\footnote{For $D = 5$, this is consistent with equation (3.3) of \cite{chlupo} after correction of typographical errors, namely changing signs there to give terms $+ a^2 g^2 [12 \sin^2 \gq - s^2 (17 - 16 \sin^2 \gq) - b^4 g^4 s^2 (1 - 8 \sin^2 \gq)]$.}
\bea
C{^t}{_{r t r}} & = & \fr{(D - 3) m}{\Xi g^2 r^{D + 1}} \bigg[ [D - 2 + a^2 g^2 - (D - 1) g^2 y^2] (1 + s_1^2 + s_2^2 + a^2 g^2 s_1^2 s_2^2) \nnr
&& + (s_1^2 + s_2^2) \bigg( -1 + g^2 y^2 + a^4 g^4 - a^2 g^4 y^2 + \fr{a^2 g^2 + g^2 y^2 - a^4 g^4 - a^2 g^4 y^2}{D - 2} \bigg) \bigg] \nnr
&& + \cO \bigg( \fr{1}{r^{D + 2}} \bigg) .
\eea

In summary, we find the thermodynamic quantities for $D = 4, 5, 6, 7$
\bea
E & = & \fr{\cA_{D - 2} m}{4 \pi \Xi} \bigg[ \bigg( \fr{1}{\Xi} + \fr{D - 4}{2} \bigg) (1 + a^2 g^2 s_1^2 s_2^2) + \bigg( \fr{1}{\Xi} + \fr{D - 5}{2} \bigg) (s_1^2 + s_2^2) \bigg] , \nnr
S & = & \fr{\cA_{D - 2} r_+^{D - 4} \sr{(r_+^2 + a^2 + 2 m s_1^2 / r_+^{D - 5}) (r_+^2 + a^2 + 2 m s_2^2 / r_+^{D - 5})}}{4 \Xi} , \nnr
T & = & \fr{\gD_r ' (r_+)}{2 \pi \sr{(r_+^2 + a^2 + 2 m s_1^2 / r_+^{D - 5}) (r_+^2 + a^2 + 2 m s_2^2 / r_+^{D - 5})}} , \nnr
J & = & \fr{\cA_{D - 2} m a c_1 c_2 \wtd{c}_1 \wtd{c}_2}{4 \pi \Xi^2} , \qd \gW = \fr{a r_+^{D - 5} [1 + g^2 (r_+^2 + 2 m s_1^2 / r_+^{D - 5})] [1 + g^2 (r_+^2 + 2 m s_2^2 / r_+^{D - 5})]}{2 m c_1 c_2 \wtd{c}_1 \wtd{c}_2} , \nnr
Q_1 & = & \fr{(D - 3) \cA_{D - 2} m s_1 c_1 \wtd{c}_2}{8 \pi \Xi} , \qd \gF_1 = \fr{2 m s_1 c_1 \wtd{c}_2}{r_+^{D - 5} (r_+^2 + a^2 + 2 m s_1^2 / r_+^{D - 5})} , \nnr
Q_2 & = & \fr{(D - 3) \cA_{D - 2} m s_2 c_2 \wtd{c}_1}{8 \pi \Xi} , \qd \gF_2 = \fr{2 m s_2 c_2 \wtd{c}_1}{r_+^{D - 5} (r_+^2 + a^2 + 2 m s_2^2 / r_+^{D - 5})} ,
\eea
where
\ben
\cA_{D - 2} = \fr{2 \pi^{(D - 1)/2}}{\gC ((D - 1) / 2)}
\een
is the volume of the unit sphere $S^{D - 2}$.  Note the particular values $\cA_2 = 4 \pi$, $\cA_3 = 2 \pi^2$, $\cA_4 = \tf{8}{3} \pi^2$ and $\cA_5 = \pi^3$.


\sse{Supersymmetric solutions}


From considering eigenvalues of the Bogomolny matrix, the BPS condition for the single-rotation solutions, up to a choice of signs, is \cite{cvgilupo}
\ben
E - g J - Q_1 - Q_2 = 0 .
\een
The parameters of a BPS solution are constrained by
\bea
&& c_1^2 \wtd{c}_2^2 - (1 - a^2 g^2) (s_1 c_1 \wtd{c}_2 + s_2 c_2 \wtd{c}_1) - 2 a g c_1 c_2 \wtd{c}_1 \wtd{c}_2 \nnr
&& + (D - 4) (1 - a^2 g^2) (1 + s_1^2 + s_2^2 + a^2 g^2 s_1^2 s_2^2 - s_1 c_1 \wtd{c}_2 - s_2 c_2 \wtd{c}_1) = 0 .
\eea
This is solved by $a g s_1 s_2 = 1$, with $\gd_1$ and $\gd_2$ positive.  Then the radial function of the metric becomes
\ben
\gD_r = \bigg( \fr{2 m s_1 s_2 g}{r^{D - 5}} - \fr{1}{g s_1 s_2} \bigg) ^2 + g^2 r^4 + 2 m (s_1^2 + s_2^2) g^2 r^{7 - D} + \bigg( 1 + \fr{1}{s_1^2 s_2^2} \bigg) r^2 .
\een
The first term might vanish, but the remaining terms are manifestly positive for positive $r$.  Since the radial function cannot have any positive roots, there are no supersymmetric AdS black holes amongst these single-rotation 2-charge solutions.


\sse{Hidden symmetry}


We now study hidden symmetries of our metrics, exhibiting Killing tensors and studying the separability of the Hamilton--Jacobi equation for geodesic motion and the Klein--Gordon equation.  These properties have previously been studied for a different class of warped sphere generalizations of the single-rotation Kerr--AdS metrics in \cite{chgilupo}.


\ssse{Killing tensor}


The string frame metric $\df \wtd{s}^2$ is related to the original Einstein frame metric $\ds$ by $\ds = (H_1 H_2)^{1 / (D - 2)} \df \wtd{s}^2$.  The inverse string frame metric is
\bea
\dsi & = & - \fr{a^4}{r^2 + y^2} \bigg( \fr{\wtd{V}_r^2}{\gD_r} - \fr{\wtd{V}_y^2}{\gD_y} \bigg) \, \pd_t^2 - \fr{2 m c_1 c_2 \wtd{c}_1 \wtd{c}_2 a}{(r^2 + y^2) r^{D - 5} \gD_r} \, 2 \, \pd_t \, \pd_\gf + \fr{1}{r^2 + y^2} \bigg( \fr{V_y^2}{\gD_y} - \fr{V_r^2}{\gD_r} \bigg) \, a^2 \, \pd_\gf^2 \nnr
&& + \fr{\gD_r}{r^2 + y^2} \, \pd_r^2 + \fr{\gD_y}{r^2 + y^2} \, \pd_y^2 + \fr{a^2}{r^2 y^2} \, \pd \gW_{D - 4}^2 ,
\eea
where $\pd \gW_{D - 4}^2$ is the inverse of $\df \gW_{D - 4}^2$.

A rank-2 Killing--St\"{a}ckel tensor is a symmetric tensor $K_{a b}$ that satisfies $\grad_{( a} K_{b c )} = 0$.  More generally, a symmetric tensor $Q_{a b}$ that satisfies an equation of the form $\grad_{( a} Q_{b c)} = q_{( a} g_{b c)}$ for some $q_a$ is called a rank-2 conformal Killing--St\"{a}ckel tensor.  In $D$ dimensions, we can express $q_a$ in terms of $Q_{a b}$ as $q_a = \tf{1}{D + 2} (\pd_a Q{^b}{_b} + 2 \grad_b Q{^b}{_a})$.

One can read off a rank-2 Killing--St\"{a}ckel tensor for the string frame metric:
\ben
\wtd{K}^{a b} \, \pd_a \, \pd_b = \fr{a^4 \wtd{V}_y^2}{\gD_y} \, \pd_t^2 + \fr{V_y^2 a^2}{\gD_y} \, \pd_\gf^2 + \gD_y \, \pd_y^2 + \fr{a^2}{y^2} \, \pd \gW_{D - 4}^2 - y^2 \bigg( \fr{\pd}{\pd \wtd{s}} \bigg) ^2 .
\een
$\wtd{K}^{a b}$ is not irreducible, i.e.~is not merely a linear combination of the metric and products of Killing vectors.  There is an induced rank-2 conformal Killing--St\"{a}ckel tensor, with components $Q^{a b} = \wtd{K}^{a b}$, for any conformally related metric, in particular the Einstein frame metric.  An Einstein frame Killing--St\"{a}ckel tensor cannot be constructed for $D \geq 5$ \cite{chow5}.


\ssse{Separability}


The string frame metric possesses a rank-2 Killing–-St\"{a}ckel tensor and satisfies the properties for a separability structure, guaranteeing that string frame geodesic motion is completely integrable.  We shall demonstrate this explicitly, and also demonstrate the separability of the Einstein frame massless Klein–-Gordon equation.  Without rotation, the enhanced isometry group alone guarantees separability.

The Hamilton--Jacobi equation for geodesic motion in the string frame metric is
\ben
\fr{\pd S}{\pd \gl} + \fr{1}{2} \wtd{g}^{a b} \, \pd_a S \, \pd_b S = 0 ,
\een
where $S$ is Hamilton's principal function, $\pd_a S = p_a = \df x_a / \df \gl$, $p_a$ are momenta conjugate to $x^a$, and $\gl$ is an affine parameter.  Consider the ansatz
\ben
S = \tf{1}{2} \mu^2 \gl - E t + L \gf + S_r (r) + S_y (y) + W(x^i) ,
\een
where the constants $p_t = -E$ and $p_\gf = L$ are momenta conjugate to the ignorable coordinates $t$ and $\gf$, representing energy and angular momentum respectively, and $x^i$ are coordinates for $S^{D - 4}$.  $\mu$ is the mass of the particle, satisfying $p^a p_a = - \mu^2$.  We therefore have additive separability:
\bea
&& - \bigg( \fr{\wtd{V}_r^2}{\gD_r} - \fr{\wtd{V}_y^2}{\gD_y} \bigg) a^4 E^2 + \fr{4 m a c_1 c_2 \wtd{c}_1 \wtd{c}_2}{r^{D - 5} \gD_r} E L + \bigg( \fr{V_y^2}{\gD_y} - \fr{V_r^2}{\gD_r} \bigg) a^2 L^2 + \gD_r \bigg( \fr{\df S_r}{\df r} \bigg) ^2 + \gD_y \bigg( \fr{\df S_y}{\df y} \bigg) ^2 \nnr
&& + a^2 \bigg( \fr{1}{r^2} + \fr{1}{y^2} \bigg) C_1 + \mu^2 (r^2 + y^2) = 0 , \eea
where we have used the separation constant
\ben
C_1 = g^{i j} \, \pd_i W \, \pd_j W .
\een
Introducing another separation constant $C_2$, we have
\bea
\fr{\df r}{\df \gl} & = & \wtd{g}^{r r} p_r = \fr{\gD_r}{r^2 + y^2} \fr{\df S_r}{\df r} , \nnr
\fr{\df y}{\df \gl} & = & \wtd{g}^{y y} p_y = \fr{\gD_y}{r^2 + y^2} \fr{\df S_y}{\df y} ,
\eea
where
\bea
S_r & = & \int \! \df r \, \fr{1}{\gD_r} \sr{\wtd{V}_r^2 a^4 E^2 - \fr{4 m a c_1 c_2 \wtd{c}_1 \wtd{c}_2 E L}{r^{D - 5}} + V_r^2 a^2 L^2 - \bigg( C_2 + \fr{a^2 C_1}{r^2} + \mu^2 r^2 \bigg) \gD_r} , \nnr
S_y & = & \int \! \df y \, \fr{1}{\gD_y} \sr{- \wtd{V}_y^2 a^4 E^2 - V_y^2 a^2 L^2 + \bigg( C_2 - \fr{a^2 C_1}{y^2} - \mu^2 y^2 \bigg) \gD_y} ,
\eea
which determines $r (\gl)$ and $y (\gl)$ by quadratures.  We then have
\bea
\fr{\df t}{\df \gl} & = & \wtd{g}^{t t} p_t + \wtd{g}^{t \gf} p_\gf = \fr{a^4 E}{r^2 + y^2} \bigg( \fr{\wtd{V}_r^2}{\gD_r} - \fr{\wtd{V}_y^2}{\gD_y} \bigg) - \fr{2 m a r^{5 - D} c_1 c_2 \wtd{c}_1 \wtd{c}_2 L}{(r^2 + y^2) \gD_r} , \nnr
\fr{\df \gf}{\df \gl} & = & \wtd{g}^{t \gf} p_t + \wtd{g}^{\gf \gf} p_\gf = \fr{2 m a r^{5 - D} c_1 c_2 \wtd{c}_1 \wtd{c}_2 E}{(r^2 + y^2) \gD_r} + \fr{a^2 L}{r^2 + y^2} \bigg( \fr{V_y^2}{\gD_y} - \fr{V_r^2}{\gD_r} \bigg) ,
\eea
which determines $t (\gl)$ and $\gf (\gl)$ by quadratures.  Combined with the fact that geodesic motion on the round sphere $S^{D - 4}$ is completely integrable, it follows that geodesic motion on the full string frame metric is completely integrable.  Geodesic motion on the Einstein frame metric involves replacing $\mu^2 (r^2 + y^2)$ above by $\mu^2 (H_1 H_2)^{1/(D - 2)} (r^2 + y^2)$, which spoils separability.  If $D \geq 5$, then only for massless particles, $\mu = 0$, is Einstein frame geodesic motion completely integrable.

Now consider the minimally coupled massless Klein--Gordon equation for the Einstein frame metric, $\square \gF = 0$.  Note that
\ben
\sr{-g} = \fr{(H_1 H_2)^{1/(D - 2)} (r^2 + y^2)}{a \Xi} \bigg( \fr{r y}{a} \bigg) ^{D - 4} \sr{g_{(D - 4)}} ,
\een
where $g_{(D - 4)}$ is the determinant of the round metric on $S^{D - 4}$.  The d'Alembetian $\square = (1/\sqrt{-g}) \pd_a (\sr{-g} g^{a b} \pd_b)$ is therefore
\bea
\square & = & \fr{(H_1 H_2)^{-1/(D - 2)}}{(r^2 + y^2)} \bigg[ - a^4 \bigg( \fr{\wtd{V}_r^2}{\gD_r} - \fr{\wtd{V}_y^2}{\gD_y} \bigg) \pd_t^2 - \fr{4 m c_1 c_2 \wtd{c}_1 \wtd{c}_2 a}{r^{D - 5} \gD_r} \, \pd_t \, \pd_\gf + \bigg( \fr{V_y^2}{\gD_y} - \fr{V_r^2}{\gD_r} \bigg) \pd_\gf^2 \nnr
&& + \fr{1}{r^{D - 4}} \pd_r (r^{D - 4} \gD_r \pd_r) + \fr{1}{y^{D - 4}} \pd_y (y^{D - 4} \gD_y \pd_y) + a^2 \bigg( \fr{1}{r^2} + \fr{1}{y^2} \bigg) \nabla^2_{(D - 4)} \bigg] ,
\eea
where $\nabla^2_{(D - 4)}$ is the $S^{D - 4}$ Laplacian.  Consider the ansatz
\ben
\gF = \gF_r (r) \gF_y (y) \expe{\im (k \gf - \gw t)} \gF (x^i) ,
\een
where $\gF (x^i)$ is an eigenfunction of $\grad^2_{(D - 4)}$ with eigenvalue $\gl = \ell (\ell + D - 5)$.  Then we have the separated equations\footnote{We correct some typographical errors in \cite{chow3}.}
\bea
&& \fr{1}{r^{D - 4}} \fr{\df}{\df r} \bigg( r^{D - 4} \gD_r \fr{\df \gF_r}{\df r} \bigg) + \fr{a^4 \wtd{V}_r^2 \gw^2 - 4 c_1 c_2 \wtd{c}_1 \wtd{c}_2 a m r^{5 - D} \gw k + V_r^2 k^2}{\gD_r} \gF_r + \fr{a^2 \gl}{r^2} \gF_r = C , \nnr
&& \fr{1}{y^{D - 4}} \fr{\df}{\df y} \bigg( y^{D - 4} \gD_y \fr{\df \gF_y}{\df y} \bigg) - \fr{a^4 \wtd{V}_y^2 \gw^2 + V_y^2 k^2}{\gD_y} \gF_y + \fr{a^2 \gl}{y^2} \gF_y = -C ,
\eea
where $C$ is a separation constant.  These are generally Fuchsian second-order ordinary differential equations.


\se{2-charge static AdS black holes in arbitrary dimensions}


So far, we have considered rotating AdS black holes that are solutions of gauged supergravity theories.  For such theories to admit AdS vacua, the maximum dimension is $D = 7$.  However, without any rotation, the 2-charge static AdS black hole solutions easily generalize to arbitrary higher dimensions as solutions of a simple bosonic theory.

We generalize the $D$-dimensional Lagrangian \eq{Lagrangian} to include a scalar potential, and also specialize by taking $A_{(D - 4)} = 0$.  The Lagrangian we consider is
\ben
\cL_D = R \star 1 - \fr{1}{2} \sum_{i = 1}^2 \star \df \gvf_i \wedge \df \gvf_i - \fr{1}{2} \sum_{I = 1}^2 X_I^{-2} \star F_{(2)}^I \wedge F_{(2)}^I - V \star 1 ,
\een
again with the scalar combinations
\ben
X_1 = \expe{- \gvf_1 / \sr{2 (D - 2)} - \gvf_2 / \sr{2}} , \qd X_2 = \expe{- \gvf_1 / \sr{2 (D - 2)} + \gvf_2 / \sr{2}} ,
\een
and $F_{(2)}^I = \df A_{(1)}^I$.  The potential is
\ben
V = - g^2 [(D - 3)^2 X_1 X_2 + 2 (D - 3) (X_1 X_2)^{- (D - 3)/2} (X_1 + X_2) - (D - 5) (X_1 X_2)^{- (D - 3)}] .
\een
There are critical points of the potential at $X_1 = X_2 = 1$, which is a maximum, and at $X_1 = X_2 = [(D - 5)/(D - 3)]^{1/(D - 2)}$, which is a saddle point.  The resulting field equations are
\bea
G_{a b} & = & \sum_{i = 1}^2 \bigg( \fr{1}{2} \grad_a \gvf_i \, \grad_b \gvf_i - \fr{1}{4} \grad^c \gvf_i \, \grad_c \gvf_i \, g_{a b} \bigg) + \sum_{I = 2}^2 X_I^{-2} \bigg( \fr{1}{2} F{^I}{_a}{^c} F{^I}{_{b c}} - \fr{1}{8} F^{I c d} F{^I}{_{c d}} g_{a b} \bigg) \nnr
&& - \fr{1}{2} V g_{a b} ,
\eea
and
\bea
&& \square \gvf_1 = \fr{1}{2 \sr{2 (D - 2)}} \sum_{I = 1}^2 X_I^{-2} F^{I a b} F{^I}{_{a b}} + \fr{\sr{2} (D - 3)}{\sr{D - 2}} g^2 [ (D - 3) X_1 X_2 \nnr
&& \qqd \qqd - (D - 4) (X_1 X_2)^{- (D - 3)/2} (X_1 + X_2) + (D - 5) (X_1 X_2)^{- (D - 3)} ] , \nnr
&& \square \gvf_2 = \fr{1}{2 \sr{2}} (X_1^{-2} F^{1 a b} F{^1}{_{a b}} - X_2^{-2} F^{2 a b} F{^2}{_{a b}}) + \sr{2} (D - 3) g^2 (X_1 X_2)^{- (D - 3)/2} (X_1 - X_2) , \nnr
&& \df (X_I^{-2} \star F_{(2)}^I) = 0 .
\eea

These theories admit AdS vacua with $G_{a b} = (D - 1) (D - 2) g^2 g_{a b}$.  If $D = 4, 5, 6, 7$, then these theories are consistent bosonic truncations of gauged supergravity theories.  If we truncate so that $A_{(1)}^2 = 0$ and $X_2 = 1$, then we reduce to a theory that has recently been shown to admit charged rotating AdS black holes \cite{wu}; Killing spinors \cite{liluwa}, from which the bosonic equations arise by projection of the corresponding integrability condition; and a pseudosymmetric extension \cite{liluwa2}, for which the theory is invariant under supersymmetry-type transformations up to quadratic order in fermions.

A 2-charge static AdS black hole solution is
\bea
\ds & = & (H_1 H_2)^{1/(D - 2)} \bigg( - \fr{f}{H_1 H_2} \, \df t^2 + \fr{\df r^2}{f} + r^2 \, \df \gW_{D - 2}^2 \bigg) , \nnr
A_{(1)}^I & = & \fr{2 m s_I c_I}{H_I r^{D - 3}} \, \df t , \nnr
X_1 & = & H_1^{-(D - 1)/2(D - 2)} H_2^{(D - 3)/2 (D - 2)} , \qd X_2 = H_1^{(D - 3)/2(D - 2)} H_2^{- (D - 1)/2 (D - 2)} ,
\eea
where
\ben
f = 1 - \fr{2 m}{r^{D - 3}} + H_1 H_2 g^2 r^2 , \qd H_I = 1 + \fr{2 m s_I^2}{r^{D - 3}} , \qd s_I = \sinh \gd_I , \qd c_I = \cosh \gd_I ,
\een
and $\df \gW_{D - 2}^2$ is the round metric on the unit sphere $S^{D - 2}$.

When $D = 4, 5, 7$, this reduces to previously known solutions of gauged supergravity \cite{dufliu, becvsa, cvegub, liumin}, and also in $D = 6$ when $\gd_1 = \gd_2$ \cite{cvlupo}.  When $g = 0$, the solution reduces to the 2-charge static solution of Peet \cite{peet}, which originally used the Horowitz--Sen charge parameters $\ga$ and $\gb$.

By the previous methods, we may compute the thermodynamical quantities.  If the AMD method is used to compute the mass, then one uses
\ben
C{^t}{_{r t r}} = \fr{(D - 3) m [D - 2 + (D - 3) (s_1^2 + s_2^2)]}{g^2 r^{D + 1}} + \cO \bigg( \fr{1}{r^{D + 2}} \bigg) .
\een
Alternatively, the mass can be obtained from the first law of black hole mechanics, $\df E = T \, \df S + \gF_1 \, \df Q_1 + \gF_2 \, \df Q_2$.  In summary, the thermodynamic quantities are
\bea
E & = & \fr{\cA_{D - 2}}{8 \pi} m [D - 2 + (D - 3) (s_1^2 + s_2^2)] , \nnr
S & = & \fr{\cA_{D - 2}}{4} r_+^{D - 2} \sr{(1 + 2 m s_1^2 / r_+^{D - 3}) (1 + 2 m s_2^2 / r_+^{D - 3})} , \nnr
T & = & \fr{D - 3 + (D - 1) g^2 r_+^2 + 4 g^2 m (s_1^2 + s_2^2) / r_+^{D - 5} - 4 (D - 5) g^2 m^2 s_1^2 s_2^2 / r_+^{2 D - 8}}{4 \pi r_+ \sr{(1 + 2 m s_1^2 / r_+^{D - 3}) (1 + 2 m s_2^2 / r_+^{D - 3})}} , \nnr
Q_I & = & \fr{(D - 3) \cA_{D - 2}}{8 \pi} m s_I c_I , \qd \gF_I = \fr{2 m s_I c_I}{r_+^{D - 3} + 2 m s_I^2} .
\eea


\se{Conclusion}


We have constructed solutions in $D = 5, 6, 7$ that represent AdS black holes with rotation in a single 2-plane and 2 independent $\uU (1)$ charges.  The construction is analogous to the $D = 4$ solution \cite{chow3}.  These new solutions should provide insight into the construction of more general solutions of these theories that are yet to be discovered.  We have studied various properties of the solutions in a unified manner, such as thermodynamics and hidden symmetries, finding straightforward generalizations of the known results in $D = 4$.

Because of the similar structure to the solutions in different dimensions, one might wonder about the possibility of generalizing these rotating solutions to some simple bosonic theory in any dimension.  However, there are substantial differences between the theories in $D = 4, 5, 6, 7$, so it is unclear if this is possible.  Without rotation, we have constructed static solutions in any dimension $D \geq 4$ that represent non-rotating AdS black holes with 2 independent $\uU (1)$ charges.  Although these static solutions do not fit into any gauged supergravity theory for $D \geq 8$, they may still be of use in understanding the AdS/CFT correspondence.




\end{document}